\documentclass[11pt]{article}
\usepackage{graphicx,amsmath}

\begin{document}

{\center{GEOMETRICAL\ SCALING IN HIGH ENERGY HADRONIC 
COLLISIONS\footnote{Presented at QCD Moriond 2012}\\~\\
MICHAL PRASZALOWICZ \\~\\
M. Smoluchowski Inst. of Physics, Jagellonian University\\ Krakow, Poland\\}}

\vspace{1cm}

\noindent{After introducing the concept of geometrical scaling (GS) on the
example of deep inelastic ep scattering, we show that GS is also present in the $p_{\rm T}$
spectra measured by  the LHC. We discuss simple phenomenological signatures of GS and
its applications.}

\vspace{1cm}


It is known that gluonic parton density rapidly increases  at low Bjorken $x$
\cite{HERAcombined}.
Such growth has to be tamed at some point. The scale at which this happens is
called saturation scale $Q_{\mathrm{{s}}}(x)$ and it depends on the Bjorken $x$.
The explicit form of the saturation scale follows from the fact that
$Q_{\mathrm{{s}}}^{2}(x)$ is related to the gluon distribution in the proton
at low $x$ \cite{GolecBiernat:1998js}:%
\begin{equation}
Q_{\mathrm{{s}}}^{2}(x)=Q_{0}^{2}\left(  {x}/{x_{0}}\right)  ^{-\lambda
}\label{defQsat}%
\end{equation}
where $Q_{0}\sim1$ GeV and $x_{0}\sim10^{-3}$ are free parameters whose
precise values can be extracted from fits to the HERA data. Power $\lambda$ is
known to be of the order $\lambda\sim0.2\div0.3$ and Bjorken $x$ is defined
as
\begin{equation}
x=Q^{2}/{(Q^{2}+W^{2}-M_{p}^{2})}\label{Bjx}%
\end{equation}
where $M_{p}$ stands for the proton mass.

Geometrical scaling \cite{Stasto:2000er} consists in the fact that
for sufficiently low $x$ the reduced
$\gamma^{\ast}p$ cross section $\sigma_{\gamma^{\ast}p}(W,Q^{2})\sim
F_{2}(x,Q^{2})/Q^{2}$ depends in fact only on the
scaling variable $\tau$:
\begin{equation}
\sigma_{\gamma^{\ast}p}=\text{\textrm{function}}(\tau),\;\text{where}%
\;\tau=Q^{2}/Q_{\text{s}}^{2}(x).\label{GS1}%
\end{equation}
This is depicted in Fig.1. where the combined HERA data \cite{HERAcombined}
for different scattering energies $W$ are plotted in terms of $Q^{2}$(left)
and in terms of $\tau$ (right). Quantitative analysis of the combined HERA data and
the details of the $W$ binning will be presented elsewhere \cite{MPTS}.

\begin{figure}[t!]
\centering
\includegraphics[scale=0.75]{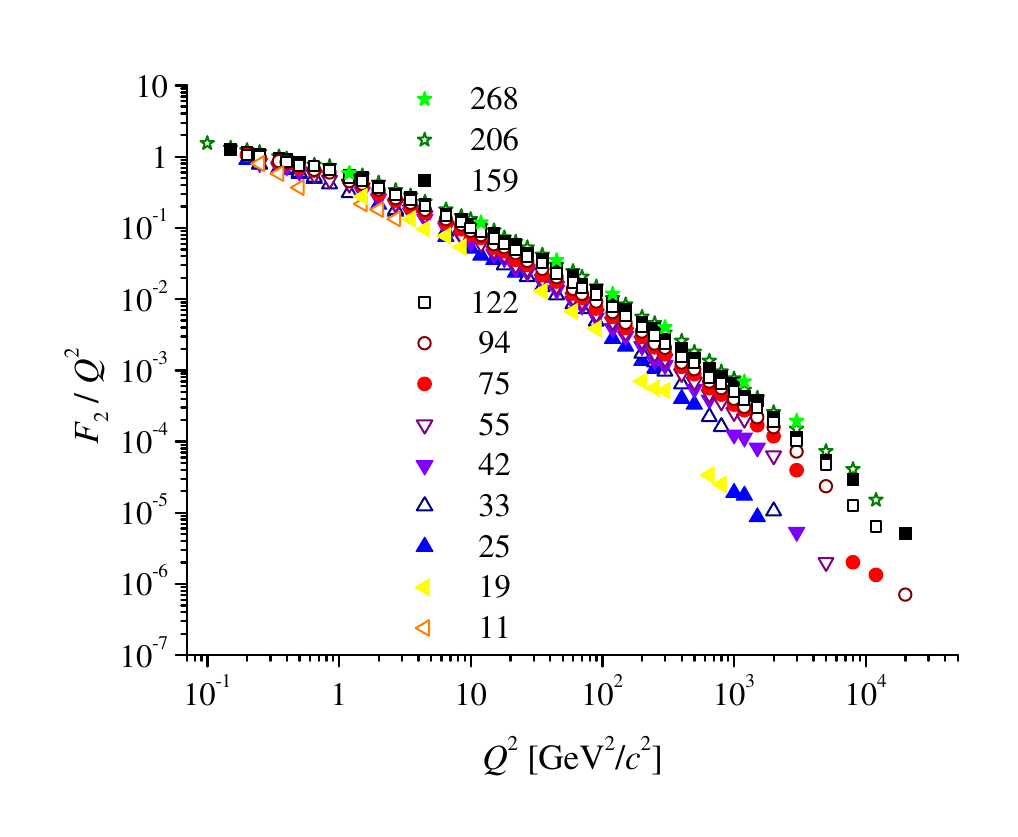}\;
\includegraphics[scale=0.75]{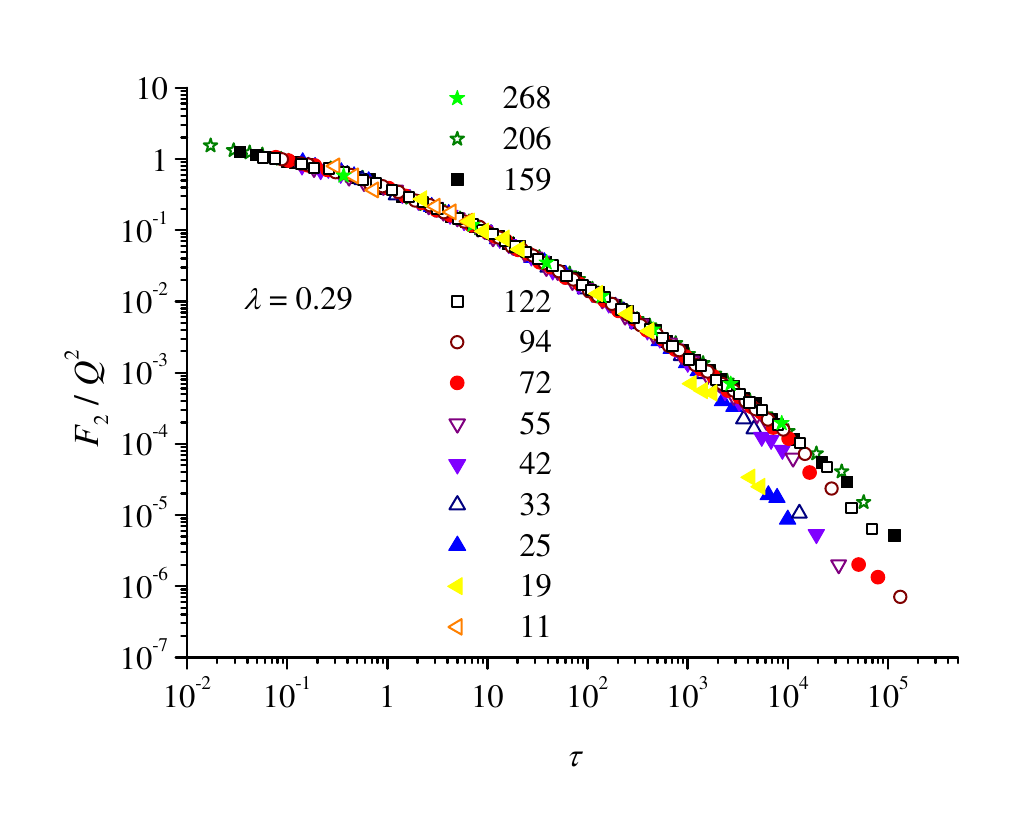}
\caption{Geometrical scaling in DIS.}%
\label{fig:DIS}
\end{figure}



In pp collisions particles of low and moderate $p_{\mathrm{{T}}}$  (and given
rapidity $y$) are produced mainly from scattering of gluons carrying
longitudinal momentum fractions $x_{1,2}$:
\begin{equation}
x_{1,2}=e^{\pm y}\,{p_{\mathrm{{T}}}}/{W}\;\;\;\mathrm{with}\;\;\;W=\sqrt
{s}.\label{x12}%
\end{equation}

If gluonic densities in pp collisions are characterized by the saturation
scale (\ref{defQsat}), then also $dN/d\eta d^{2}p_{\mathrm{{T}}}$ should
scale. Therefore geometrical scaling for the multiplicity distribution in pp
collisions \cite{McLerran:2010ex,Praszalowicz:2011tc} states that particle
spectra depend on the scaling variable%
\begin{equation}
\tau=p_{\mathrm{{T}}}^{2}/{Q_{\mathrm{{s}}}^{2}(p_{\mathrm{{T}}}%
,W)}\label{taudef}%
\end{equation}
where $Q_{\mathrm{{s}}}^{2}(p_{\mathrm{{T}}},W)$ is the saturation scale
(\ref{defQsat}) at $x_{1}\sim x_{2}$ (\ref{x12}):
\begin{equation}
Q_{\mathrm{{s}}}^{2}(p_{\mathrm{{T}}},W)=Q_{0}^{2}\left(  p_{\mathrm{{T}}%
}/{(W\times10^{-3})}\right)  ^{-\lambda}\label{Qsdef}%
\end{equation}
where we have neglected rapidity dependence of $x_{1,2}$. Factor $10^{-3}$
corresponds to the choice of the energy scale (arbitrary at this moment
$x_{0}$ in Eq.(\ref{defQsat})). Hence
\begin{equation}
N_{\mathrm{{ch}}}(W,p_{\mathrm{{T}}}){=}\left.  \frac{dN_{\mathrm{{ch}}}%
}{d\eta d^{2}p_{\mathrm{{T}}}}\right\vert _{W}=\frac{1}{Q_{0}^{2}}%
F(\tau)\label{GSpp}%
\end{equation}
with $Q_{0}\sim1$ GeV. Here $F(\tau)$ is a universal function of $\tau$. This
is depicted in Fig. 2 where the $p_{\rm T}$ spectra measured by CMS
\cite{Khachatryan:2010xs} ale plotted in terms of $p_{\rm T}^2$ (left) and
$\tau$ (right).

\begin{figure}[b!]
\centering
\includegraphics[scale=0.9]{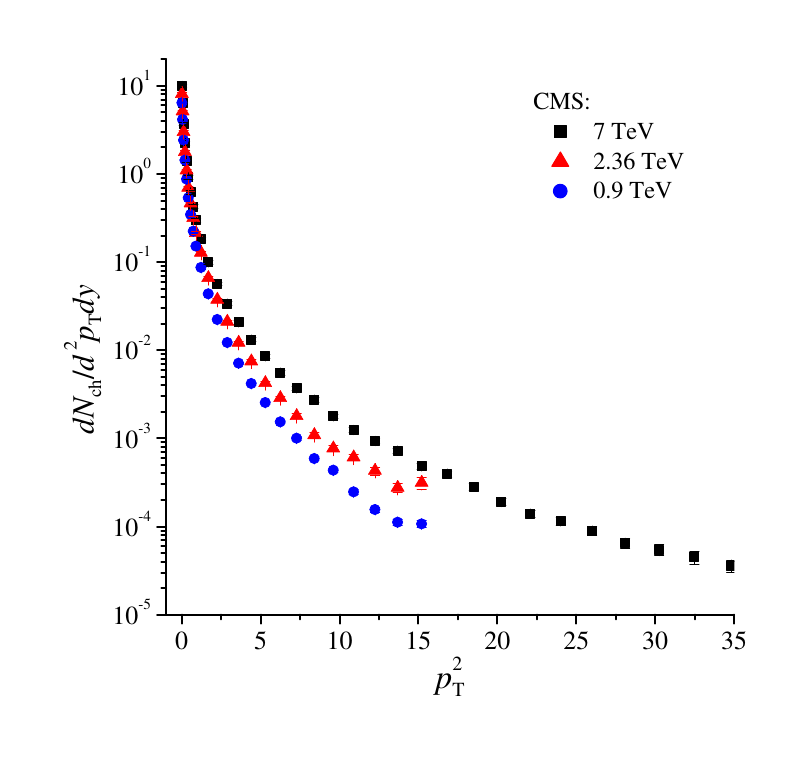}\;
\includegraphics[scale=0.9]{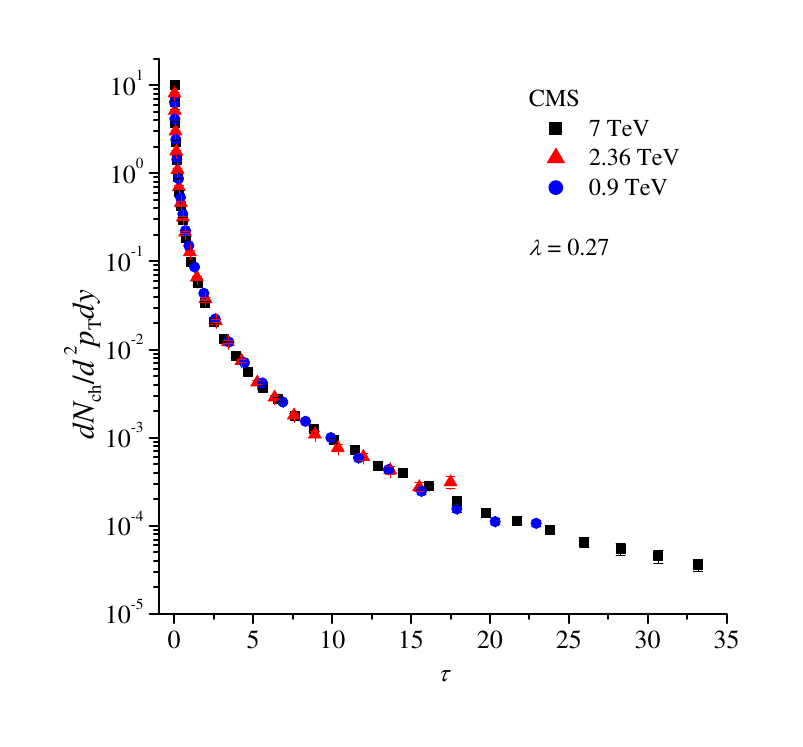}
\caption{Geometrical scaling in pp.}%
\label{fig:DIS}
\end{figure}

In order to examine the quality of geometrical scaling in pp collisions in
Ref.[8] \nocite{Praszalowicz:2011rm} we have considered ratios $R_{W_{1}/W_{2}}$
\begin{equation}
R_{W_{1}/W_{2}}(p_{\mathrm{{T}}}){=} \frac{N_{\mathrm{{ch}}}(W_{1}%
,p_{\mathrm{{T}}})}{N_{\mathrm{{ch}}}(W_{2},p_{\mathrm{{T}}})}.
\end{equation}

Here, following Ref.[9]\nocite{Praszalowicz:2011jf}  we shall discuss another way
of establishing
geometrical scaling, at least qualitatively. Note that if at two different
energies $W_{1}$ and $W_{2}$ multiplicity distributions are equal
\begin{equation}
N_{\mathrm{{ch}}}(W_{1},{p_{\,\mathrm{{T}}}^{(1)}})=N_{\mathrm{{ch}}}%
(W_{2},{p_{\,\mathrm{{T}}}^{(2)}})\label{Nchequality}%
\end{equation}
then this means that they correspond to the same value of variable $\tau$
(\ref{taudef}). As a consequence%
\begin{equation}
p_{\,\mathrm{{T}}}^{(1)\,2}\left(  {p_{\,\mathrm{{T}}}^{(1)}}/{W_{1}}\right)
^{\lambda}=p_{\,\mathrm{{T}}}^{(2)\,2}\left(  {p_{\,\mathrm{{T}}}^{(2)}%
}/{W_{2}}\right)  ^{\lambda}\label{pTW}%
\end{equation}
for constant $\lambda$. Equation (\ref{pTW}) implies%
\begin{equation}
S_{W_{1}/W_{2}}^{p_{\mathrm{{T}}}}{=}{p_{\,\mathrm{{T}}}^{(1)}}%
/{p_{\,\mathrm{{T}}}^{(2)}}=\left(  {W_{1}}/{W_{2}}\right)  ^{\frac{\lambda
}{2+\lambda}}.\label{ratiospT}%
\end{equation}

\begin{figure}[h]
\centering
\includegraphics[scale=0.7]{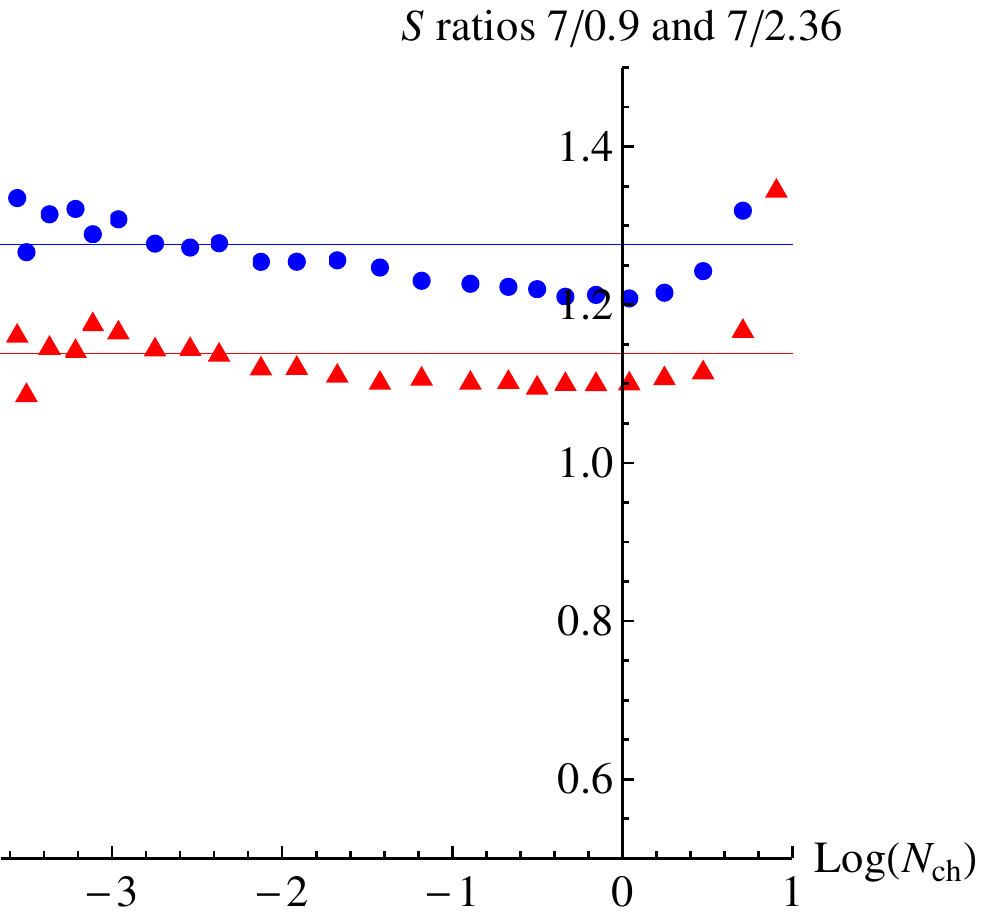}
\caption{
$S^{p_{\rm{T}}}_{W_{1}/W_{2}}$ ratios for CMS pp spectra}
\label{fig:pTratios}%
\end{figure}

Ratios $S^{p_{\mathrm{{T}}}}_{W_{1}/W_{2}}$ for pp non-single diffractive
spectra measured by the CMS \cite{Khachatryan:2010xs} collaboration at the LHC
are plotted in  Fig.~\ref{fig:pTratios} together with the
straight horizontal lines corresponding to the r.h.s. of Eq.(\ref{ratiospT})
for $\lambda=0.27$. We see approximate constancy of $S_{W_{1}/W_{2}%
}^{p_{\mathrm{{T}}}}$ over the wide range of $N_{\mathrm{{ch}}}$. A small rise
of $S_{W_{1}/W_{2}}^{p_{\mathrm{{T}}}}$ with decreasing $N_{\mathrm{{ch}}}$
corresponds to the residual $p_{\mathrm{{T}}}$-dependence~\cite{Praszalowicz:2011tc}
of the exponent
$\lambda$.

This is the simplest way of looking for GS in
the $p_{\mathrm{{T}}}$ spectra.  An obvious advantage is
that it is very easy to do. An obvious disadvantage consists in the
fact that it is difficult to attribute sensible error to the ratios
$S_{W_{1}/W_{2}}^{p_{\mathrm{{T}}}}$, so for quantitative purposes it is
better to consider ratios $R_{W_{1}/W_{2}}$.

One of the immediate applications of GS is its ability to \emph{predict}
$p_{\text{T}}$ spectra at yet unmeasured energies. This was a crucial problem
in calculating the so called nuclear modification factor $R_{AA}$ for Pb-Pb
collisions at the LHC. $R_{AA}$ is essentially a ratio of nuclear to pp
spectra at the same scattering energy normalized by the number of binary
collisions. In the first heavy ion LHC run the c.m.s. energy per nucleon was
2.76 GeV and there was no data for pp collisions at this energy until the late
run in 2011. In Fig.~4 we plot $R_{AA}$ as published by ALICE~\cite{Aamodt:2010jd}.
Black (upper) stars correspond the 2010 data where proton spectrum has been 
interpolated from the measurements at other energies (two solid grey lines 
correspond to the estimated
uncertainty). Pink (lower) stars in turn correspond to the preliminary data where pp
spectrum has been measured in a dedicated 2.76 GeV run~\cite{Knichel}.
Triangles and circles correspond to our rough estimate of $R_{AA}$ where
ALICE measured Pb-Pb spectrum~\cite{Aamodt:2010jd} has been divided by
the theoretical pp spectrum obtained from the hypothesis of geometrical scaling
for two different values of $\lambda$.

\begin{figure}[h]
\centering
\includegraphics[scale=0.80]{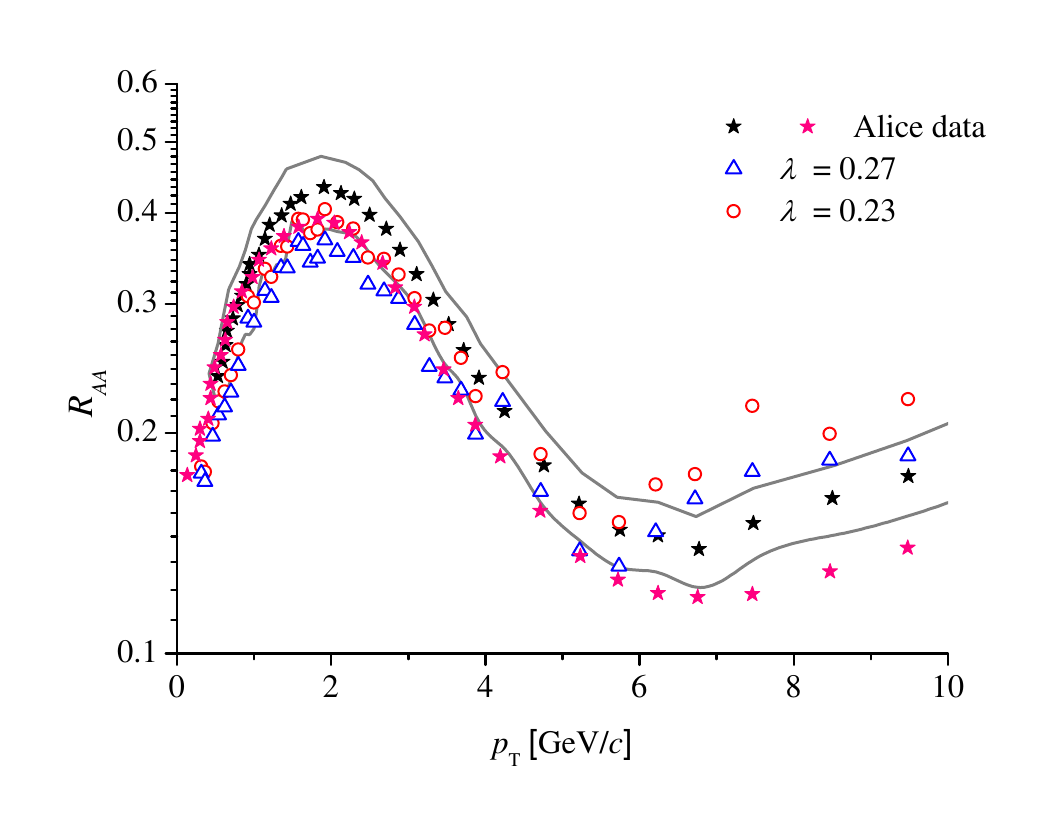}\;
\caption{$R_{AA}$ measured by ALICE (stars) and GS prediction (triangles and circles, see text) .}%
\label{fig:RAA}
\end{figure}


In this talk we have argued that geometrical scaling is a universal phenomenon
observed in DIS and in pp scattering at the LHC (for theoretical background
see lectures by L. McLerran~\cite{McLerran:2010ub}). After illustrating how GS
works in these two processes we have proposed a simple procedure to look for
geometrical scaling in the $p_{\mathrm{T}}$ spectra, namely to construct
ratios of transverse momenta corresponding to the same multiplicity. We have
used GS to predict the $p_{\text{T}}$ spectra at yet unmeasured energies.

Many aspects of GS require further studies. Firstly, new data at higher
energies (to come) have to be examined. Secondly, more detailed analysis
including identified particles and rapidity dependence has to be performed. On
theoretical side the universal shape $F(\tau)$ has to be found and its
connection to the unintegrated gluon distribution has to be studied. That will
finally lead to perhaps the most difficult part, namely to the breaking of GS
in pp.

\section*{Acknowledgements}

The author wants to thank Larry McLerran and Andrzej Bialas for stimulating
discussions. This work was supported in part by the Polish NCN grant 2011/01/B/ST2/00492.



\begin{thebibliography}{99}                                                                                               %
\bibitem {HERAcombined} F.~D.~Aaron \textit{et al.} [H1 and ZEUS
Collaboration],
JHEP \textbf{1001} (2010) 109.




\bibitem {GolecBiernat:1998js} K.J.~Golec-Biernat, M.~W\"usthoff,
Phys.\ Rev.\ D \textbf{59} (1998) 014017,
D \textbf{60} (1999) 114023.




\bibitem {Stasto:2000er}A.M.~Stasto, K.J.~Golec-Biernat, J.~Kwiecinski,
Phys.\ Rev.\ Lett.\ \textbf{86} (2001) 596.




\bibitem {MPTS}
M. Praszalowicz, T. Stebel, in preperation.


\bibitem {McLerran:2010ex}L.~McLerran, M.~Praszalowicz,
Acta Phys.\ Pol.\ B \textbf{41} (2010) 1917
and \textbf{42} (2011) 99.


\bibitem {Praszalowicz:2011tc}M.~Praszalowicz,
Phys. Rev. Lett. 106 (2011) 142002.



\bibitem {Khachatryan:2010xs}V.~Khachatryan \textit{et al.} [CMS
Collaboration],
JHEP \textbf{1002} (2010) 041,
Phys.\ Rev.\ Lett.\ \textbf{105} (2010) 022002,
JHEP \textbf{1101} (2011) 079.



\bibitem {Praszalowicz:2011rm}M.~Praszalowicz,
Acta Phys.\ Pol.\ \textbf{B42 } (2011) 1557.

\bibitem{Praszalowicz:2011jf}
  M.~Praszalowicz,
  arXiv:1112.0997 [hep-ph].


\bibitem{Aamodt:2010jd}
  K.~Aamodt {\it et al.}  [ALICE Collaboration],
  Phys.\ Lett.\ B {\bf 696} (2011) 30.


\bibitem{Knichel}
For pp spectra at 2.76 GeV see talk of M.L.  Knichel, this proceedings.



\bibitem {McLerran:2010ub}L.~McLerran,
Acta Phys.\ Pol.\ B \textbf{41} (2010) 2799.

\end{thebibliography}
\end{document}